\documentclass[twocolumn,showpacs,preprintnumbers,amsmath,amssymb]{revtex4}
\usepackage{graphicx}% Include figure files
\usepackage{dcolumn}% Align table columns on decimal point
\usepackage{bm}% bold math

\newcommand{\N}{{\mathbb{N}}}

\newcommand{\evt}{C}
\newcommand{\minf}{I}

\newcommand{\sk}{s}
\newcommand{\rsk}{S}
\newcommand{\s}{s}
\newcommand{\g}{g}
\newcommand{\rg}{G}
\newcommand{\dfn}{\equiv}
\newcommand{\tr}{{\mathrm{Tr}}\hspace{1pt}}
\newcommand{\pdf}{B_{1}}
\newcommand{\kld}{D}
\newcommand{\tdis}{d_{T}}

\newcommand{\vdis}{d_{V}}

\newcommand{\fid}{F}

\newcommand{\ren}{R}
\newcommand{\rna}{\ren_{\bbsa}}

\newcommand{\rnamp}{\ren_{\bbsa-}\upcm}
\newcommand{\rnalp}{\ren_{\bbsa-}\upcl}
\newcommand{\rne}{\ren_{E}\upky}

\newcommand{\reny}{R\'{e}nyi}

\newcommand{\seb}{h}
\newcommand{\cmm}{,}
\newcommand{\unt}{U}
\newcommand{\untd}{\unt^{\dag}}
\newcommand{\sbd}{{\cal{X}}}
\newcommand{\bsa}{a}

\newcommand{\bsab}{\bar{\bsa}}
\newcommand{\bsai}{\bsa_{\ii}}
\newcommand{\bbsa}{\bsa}
\newcommand{\bbsat}{\tilde{\bbsa}}

\newcommand{\bbsab}{\bar{\bbsa}}
\newcommand{\bsb}{b}
\newcommand{\bbsb}{\bsb}

\newcommand{\dta}{x}
\newcommand{\dtad}{\dta'}
\newcommand{\dtai}{\dta_{\ii}}
\newcommand{\bdta}{\dta}
\newcommand{\rbdta}{X}
\newcommand{\dtb}{y}
\newcommand{\bdtb}{\dtb}

\newcommand{\bdtbt}{\tilde{\bdtb}}
\newcommand{\rbdtbt}{\tilde{Y}}

\newcommand{\dte}{z}
\newcommand{\rdte}{Z}
\newcommand{\dtes}{\dte^{*}}
\newcommand{\st}{\rho}

\newcommand{\stz}{\st_{\bsz}}
\newcommand{\sto}{\st_{\bso}}
\newcommand{\sta}{\st_{\bsa}}
\newcommand{\stabz}{\sta\brz}
\newcommand{\stabo}{\sta\bro}

\newcommand{\hsta}{\hst_{\bsa}}
\newcommand{\stal}{\st_{A}}
\newcommand{\bst}{\bar{\st}}
\newcommand{\hst}{\hat{\st}}
\newcommand{\hstad}{\hst\ad}
\newcommand{\hstabd}{\hst\abd}
\newcommand{\hstam}{\bst_{\bbsa}}
\newcommand{\hstzm}{\bst_{\bsz}}
\newcommand{\hstom}{\bst_{\bso}}
\newcommand{\hstabm}{\bst_{\bbsab}}

\newcommand{\stad}{\st_{\bbsa\cmm\bdta}}

\newcommand{\apaty}{\apm_{\bdtbt}}
\newcommand{\ad}{_{\bbsa\cmm\dta}}
\newcommand{\abd}{_{\bbsab\cmm\dta}}

\newcommand{\oad}{_{\bsa\cmm\dta}}
\newcommand{\oadb}{_{\bsa\cmm\bar{\dta}}}
\newcommand{\oadp}{_{\bsa\cmm\dta'}}
\newcommand{\oadi}{_{\bsai\cmm\dtai}}
\newcommand{\oaz}{_{\bsa\cmm\dtz}}
\newcommand{\oao}{_{\bsa\cmm\dto}}
\newcommand{\ozd}{_{\bsz\cmm\dta}}
\newcommand{\ood}{_{\bso\cmm\dta}}
\newcommand{\oal}{_{\alpha}}
\newcommand{\ozz}{_{0\cmm0}}
\newcommand{\ozo}{_{0\cmm1}}
\newcommand{\ooz}{_{1\cmm0}}
\newcommand{\ooo}{_{1\cmm1}}
\newcommand{\brz}{^{(0)}}
\newcommand{\bro}{^{(1)}}
\newcommand{\brd}{^{(d)}}
\newcommand{\brom}{^{(1)\cm}}
\newcommand{\ostad}{\st\oad}
\newcommand{\ostzd}{\st\ozd}
\newcommand{\ostod}{\st\ood}
\newcommand{\ostal}{\st\oal}
\newcommand{\ostalz}{\ostal\brz}

\newcommand{\ostadz}{\ostad\brz}
\newcommand{\ostado}{\ostad\bro}
\newcommand{\ostzdo}{\ostzd\bro}
\newcommand{\ostodo}{\ostod\bro}
\newcommand{\apst}{\sigma}
\newcommand{\apad}{\apst\ad}
\newcommand{\apabd}{\apst\abd}
\newcommand{\apm}{\bar{\apst}}
\newcommand{\apam}{\apm_{\bbsa}}
\newcommand{\apabm}{\apm_{\bbsab}}

\newcommand{\oapad}{\apst\oad}
\newcommand{\oapadb}{\apst\oadb}
\newcommand{\oapaz}{\apst\oaz}
\newcommand{\oapao}{\apst\oao}

\newcommand{\kal}{\kk_{\alpha}}
\newcommand{\kbe}{\kk_{\beta}}
\newcommand{\kab}{\kk_{\alpha\beta}}
\newcommand{\lal}{\lambda_{\alpha}}
\newcommand{\lbe}{\lambda_{\beta}}
\newcommand{\mab}{\mu_{\alpha\beta}}
\newcommand{\maa}{\mu_{\alpha\alpha}}
\newcommand{\grm}{G}
\newcommand{\sqm}{C}
\newcommand{\la}{\langle}
\newcommand{\ra}{\rangle}
\newcommand{\eve}{E}
\newcommand{\ste}{\st_{\eve}}
\newcommand{\ms}{E}
\newcommand{\msbd}{\ms_{\bbsb\cmm\bdtb}}
\newcommand{\msad}{\ms_{\bbsa\cmm\bdtb}}

\newcommand{\mse}{\ms_{\dte}}
\newcommand{\mszz}{\ms_{\bsz\cmm\dtz}}
\newcommand{\mszo}{\ms_{\bsz\cmm\dto}}
\newcommand{\msoz}{\ms_{\bso\cmm\dtz}}
\newcommand{\msoo}{\ms_{\bso\cmm\dto}}
\newcommand{\povm}{M}
\newcommand{\povmae}{\povm_{\bbsa\cmm\bdtb\dte}}
\newcommand{\povmbae}{\povm_{\bbsab\cmm\bdtb\dte}}
\newcommand{\povmbe}{\povm_{\bdtbt\dte}}
\newcommand{\bsz}{0}
\newcommand{\bso}{1}
\newcommand{\dtz}{0}
\newcommand{\dto}{1}
\newcommand{\bin}{\{\dtz,\dto\}}
\newcommand{\stn}{N}
\newcommand{\sn}{\{1,\cdots,\stn\}}
\newcommand{\emp}{\phi}
\newcommand{\pbz}{\prob\brz}
\newcommand{\pma}{\prob_{\cm}^{\bsa}}
\newcommand{\pbad}{\prob\oad}
\newcommand{\pbadz}{\pbad\brz}
\newcommand{\pbado}{\pbad\bro}
\newcommand{\pbadd}{\pbad\brd}
\newcommand{\pbal}{\prob\oal}
\newcommand{\pbalz}{\pbal\brz}

\newcommand{\pbaz}{\prob\oaz}
\newcommand{\pbao}{\prob\oao}
\newcommand{\pbazo}{\pbaz\bro}
\newcommand{\pbaoo}{\pbao\bro}
\newcommand{\pbazd}{\pbaz\brd}
\newcommand{\pbaod}{\pbao\brd}
\newcommand{\phad}{\hat{\prob}\oad}

\newcommand{\phado}{\phad\bro}
\newcommand{\phadd}{\phad\brd}

\newcommand{\pao}{\prob_{\bsa}\bro}
\newcommand{\pam}{\prob_{\mn}^{\bsa}}

\newcommand{\pr}{{\mathrm{Pr}}}
\newcommand{\pp}{p}
\newcommand{\prob}{p}

\newcommand{\pba}{\prob_{\bbsa}}
\newcommand{\pbat}{\prob_{\bbsat}}

\newcommand{\hpa}{\prh_{\bbsa}}

\newcommand{\prh}{\hat{\prob}}

\newcommand{\praz}{\prob_{\bsz}}
\newcommand{\prao}{\prob_{\bso}}
\newcommand{\prabz}{\praz\brz}
\newcommand{\prabo}{\prao\bro}

\newcommand{\dpr}{\hpa'}
\newcommand{\as}{\ast}
\newcommand{\pas}{\prob^{\as}}
\newcommand{\prj}{P}

\newcommand{\prad}{\prj\oad}

\newcommand{\praty}{\prj_{\bdtbt}}

\newcommand{\dd}{d}

\newcommand{\cns}{c}

\newcommand{\el}{l}
\newcommand{\kk}{k}
\newcommand{\nn}{n}
\newcommand{\mm}{m}
\newcommand{\ii}{i}

\newcommand{\ch}{{\cal{H}}}
\newcommand{\chtd}{\ch_{2}}
\newcommand{\chfd}{\ch_{4}}
\newcommand{\cs}{{\cal{S}}}
\newcommand{\chn}{\ch^{\otimes\stn}}

\newcommand{\csh}{\cs(\ch)}

\newcommand{\che}{\ch_{E}}

\newcommand{\chl}{\ch^{\cl}}

\newcommand{\ds}{{\cal{D}}}
\newcommand{\co}{{\cal{C}}}
\newcommand{\ts}{{\cal{T}}}
\newcommand{\ky}{{\cal{K}}}
\newcommand{\ct}{{\cal{A}}}
\newcommand{\cl}{{\cal{L}}}
\newcommand{\cm}{{\cal{M}}}
\newcommand{\cb}{{\cal{B}}}

\newcommand{\upco}{^{\co}}
\newcommand{\upts}{^{\ts}}
\newcommand{\upky}{^{\ky}}

\newcommand{\upcm}{^{\cm}}
\newcommand{\upcl}{^{\cl}}
\newcommand{\upca}{^{\ct}}
\newcommand{\upcb}{^{\cb}}

\newcommand{\po}{\prob^{e}_{\ts}}

\newcommand{\ple}{\prob^{e}_{\cl}}
\newcommand{\mx}{+}
\newcommand{\mn}{-}

\newcommand{\plmx}{\prob^{\mx}_{\cl}}
\newcommand{\nal}{\nn^{\bsa}_{\cl}}
\newcommand{\nbzl}{\bar{\nn}^{\bsz}_{\cl}}
\newcommand{\nbol}{\bar{\nn}^{\bso}_{\cl}}
\newcommand{\nbal}{\bar{\nn}^{\bsa}_{\cl}}
\newcommand{\ner}{\nn^{e}_{\ts}}
\newcommand{\na}{\nn_{\ct}}
\newcommand{\naa}{\na^{\bsa}}
\newcommand{\nda}{\nd^{\bsa}}
\newcommand{\nka}{\nk^{\bsa}}
\newcommand{\nbd}{\nb^{\dd}}
\newcommand{\nb}{\nn_{\cb}}
\newcommand{\nd}{\nn_{\ds}}
\newcommand{\nc}{\nn_{\co}}
\newcommand{\nac}{\nc^{\bsa}}
\newcommand{\nt}{\nn_{\ts}}
\newcommand{\nk}{\nn_{\ky}}
\newcommand{\nl}{\nn_{\cl}}
\newcommand{\nm}{\nn_{\cm}}
\newcommand{\nmp}{\nm^{\bsa\mx}}
\newcommand{\nam}{\nm^{\bsa}}
\newcommand{\nzm}{\nm^{\bsz}}
\newcommand{\nom}{\nm^{\bso}}
\newcommand{\dr}{\Delta}

\newcommand{\drp}{\delta_{\prj}}
\newcommand{\dps}{\delta_{\prob}}

\newcommand{\dma}{\delta_{\cm}^{\bsa}}
\newcommand{\qq}{q}

\newcommand{\qa}{\qq_{\bsa}}

\newcommand{\qz}{\qq_{\bsz}}
\newcommand{\qo}{\qq_{\bso}}

\newcommand{\id}{I}

\newcommand{\idl}{\id_{\chl}}

\newcommand{\sucm}{\s_{\cm}}
\newcommand{\suam}{\sucm^{\bsa}}
\newcommand{\suzm}{\sucm^{\bsz}}
\newcommand{\suom}{\sucm^{\bso}}
\newcommand{\sul}{\s_{\cl}}
\newcommand{\sual}{\sul^{\bsa}}
\newcommand{\suzl}{\sul^{\bsz}}
\newcommand{\suol}{\sul^{\bso}}
\newcommand{\suml}{\sul^{m}}
\newcommand{\ep}{\epsilon}
\newcommand{\epl}{\ep_{\cl}}

\newcommand{\oml}{\omega_{\cl}}

\newcommand{\cpl}{\Pi_{\cl}}
\newcommand{\pil}{\pi_{\cl}}
\newcommand{\pilm}{\bar{\pi}_{\cl}}
\newcommand{\mul}{\mu_{\cl}}
\newcommand{\nul}{\nu_{\cl}}
\newcommand{\epet}{\ep_{\ts}^{e}}
\newcommand{\epam}{\ep_{\cm}^{\bsa}}
\newcommand{\epzm}{\ep_{\bsz}\upcm}
\newcommand{\epom}{\ep_{\bso}\upcm}
\newcommand{\eppj}{\ep^{\prj}}

\newcommand{\dsc}{T}
\newcommand{\dsad}{\dsc\oad}
\newcommand{\dsadp}{\dsc\oadp}
\newcommand{\dsaz}{\dsc\oaz}
\newcommand{\dsao}{\dsc\oao}
\newcommand{\dsap}{\dsc_{\bsa\cmm\emp}}

\begin{document}

\preprint{APS/123-QED}

\title{Security proof of
practical quantum key distribution schemes}
% Force line breaks with \\

\author{Yodai Watanabe}
\affiliation{%
National Institute of Informatics,
Research Organization of
Information and Systems\\
2-1-2 Hitotsubashi, Chiyoda-ku,
Tokyo 1018430, Japan
}%

\date{\today}% It is always \today, today,
             %  but any date may be explicitly specified

\begin{abstract}
This paper provides a security proof
of the Bennett-Brassard (BB84)
quantum key distribution protocol
in practical implementation.
To prove the security,
it is not assumed that
defects in the devices are absorbed
into an adversary's attack.
In fact, the only assumption
in the proof
is that the source is characterized.
The proof is performed by lower-bounding
adversary's {\reny} entropy
about the key before privacy amplification.
The bound reveals the leading factors
reducing the key generation rate.
\end{abstract}

% PACS, the Physics and Astronomy
 % Classification Scheme.
%\keywords{Suggested keywords}%Use showkeys class option if keyword
                              %display desired
\maketitle

One of the fundamental problems in cryptography
is to provide a way of sharing a secret random number
between two parties, Alice and Bob,
in the presence of an adversary Eve.
The quantum key distribution is
a solution to this problem\cite{be92,bb84};
indeed it allows Alice and Bob
to generate a shared secret key
securely against
Eve with unbounded resources of computation.
The security of
quantum key distribution
against general attacks
was first proved by Mayers\cite{ma01}.
Later, Shor-Preskill\cite{sp00}
provided a simple security proof %of
%quantum key distribution
based on the observation
that quantum key distribution (BB84 protocol)
is closely related to quantum error-correcting codes
(CSS codes).
Gottesmann et al.\cite{gllp04}
showed that 
the Shor-Preskill proof is still valid
as long as
the source and detector are perfect enough
so that all defects
can be absorbed into Eve's attack
(see also \cite{hm03,wmu04}
for the rate achievability of quantum codes
in the security proof).
In contrast to the security proof based on
quantum codes,
the Mayers proof
has a remarkable characteristics.
Namely in the Mayers proof,
although the source has to be
(almost) perfect,
there is no restriction on the detector;
in particular,
it can be uncharacterized.
By exchanging the role of the source and detector
in the Mayers proof,
Koashi-Preskill\cite{kp03}
provided a security proof
which applies to the case
where the detector is perfect,
but the source can be uncharacterized
(except that the averaged states
are independent of Alice's basis).
The aim of this work
is to generalize these results.
We provide a security proof of the BB84 protocol
in which the only assumption is
that the source is characterized.
In the same way as Koashi-Preskill\cite{kp03},
this can be transformed into a security proof
which is based on characteristics of the detector.
Further
we note that the security proof also applies
to the B92 protocol\cite{be92}.
%which reveals an advantage of the B92 protocol.

Let us first recall the BB84 protocol\cite{bb84}.
Let $\ch$ be a Hilbert space.
Let $\ct=\sn$, and for $\cb\subset\ct$
denote the cardinality of $\cb$
by $\nn_{\cb}$.
The BB84 protocol
is described as follows.

BB84 protocol:
(i) Alice generates two binary strings
$\bbsa\upca=\{\bsa_{\ii}\}_{\ii\in\ct}$
and $\bdta\upca=\{\dta_{\ii}\}_{\ii\in\ct}$
according to the probability distribution
%\begin{equation}
$\prob(\bbsa\upca,\bdta\upca)
=\prod_{i}\prob\oadi$.
%=\prod_{i}\pa(\bsa_{\ii},\dta_{\ii})$.
%\begin{split}
%\begin{align}
%\prob(\bbsa\upca)&=\prob(\bsa_{1})\prob(\bsa_{2})
%\cdots\prob(\bsa_{\stn}),\\
%\prob(\bdta\upca)&=\prob(\bdt_{1})\prob(\bdt_{2})
%\cdots\prob(\bdt_{\stn}),
%\end{align}
%\end{split}
%\end{equation}
(ii) Bob generates a binary string
$\bbsb\upca=\{\bsb_{\ii}\}_{\ii\in\ct}$
according to the probability distribution
$\prob(\bbsb\upca)=\prod_{\ii}\prob_{\bsb_{\ii}}$.
%$\prob(\bbsb\upca)=\prod_{\ii}\pb(\bsb_{\ii})$.
(iii) Alice sends
the quantum state on $\chn$,
%\begin{equation}
$\stad\upca=
\bigotimes_{\ii\in\ct}\st_{\bsa_{\ii}\cmm\dta_{\ii}}$,
%\end{equation}
to Bob.
(iv) Bob applies the measurement on $\chn$,
%\begin{equation}
$\{\msbd\upca\}_{\bdtb\upca}
=\big\{\bigotimes_{\ii\in\ct}
\ms_{\bsb_{\ii}\cmm\dtb_{\ii}}
\big\}_{\bdtb\upca\in\{\dtz,\dto,\emp\}^{\stn}}$,
%\end{equation}
to the received quantum state,
where $\ms_{\bsz,\emp}=\ms_{\bso,\emp}$
is the measurement corresponding to
the result that Bob cannot detect a state.
(v) Alice and Bob open $\bbsa\upca$ and $\bbsb\upca$
respectively.
Let $\ds=\{\ii\in\ct|\dtb_{\ii}\neq\emp\}$
and $\co=\{\ii\in\ds|\bsa_{\ii}=\bsb_{\ii}\}$.
%$\ts=\co\cap\rd$, $\ky=\co\backslash\rd$'¤È'¤¹'¤ë'¡¥
Alice and Bob select a random subset $\ts\subset\co$
(which does not necessarily satisfy $\nt/\nc\sim1/2$).
Let $\ky=\co-\ts$.
(vi) Alice and Bob compare $\bdta\upts$ and $\bdtb\upts$,
and count the number of errors,
$\ner=|\{\ii\in\ts|\dta_{\ii}\neq\dtb_{\ii}\}|$.
(vii) Bob estimates $\bdta\upky$
by exchanging error-correction information
with Alice.
(viii) Alice and Bob generate
a secret key $\sk$ by applying
a compression function to $\bdta\upky$.

To prove the security of the BB84 protocol,
the previous works\cite{gllp04,kp03,ma01,sp00}
assume that either Alice's source or Bob's detector
is almost perfect
in the sense that all defects in the device
can be absorbed into Eve's attack.
We wish to prove the security
of quantum key distribution
under practical implementation.
Note that
the previous security proofs have been based on
directly bounding Eve's mutual information
about the final key,
i.e. the key after privacy amplification.
In this work,
we first lower-bound Eve's {\reny} entropy
about the key before privacy amplification,
and then apply privacy amplification
in the classical information theory
which makes use of a compression function
in a universal hash family
(see \cite{bbcm95}
for the classical theory of privacy amplification).

We now provide basic definitions
%in (quantum) statistics
which will be used later
(see e.g. \cite{ha05} for details).
The variation distance between
probability distributions $\pp$ and $\qq$
is given by
$\vdis(\prob,\qq)=\frac{1}{2}
\sum_{\omega}|\prob(\omega)-\qq(\omega)|$.
%\begin{align}
%\vdis(\prob,\qq)
%&=\frac{1}{2}
%\sum_{\omega}|\prob(\omega)-\qq(\omega)|.
%\end{align}
The quantum analogue of the variation distance
is called the trace distance.
For an Hermitian operator $X$
with the spectrum decomposition $X=\sum_{i}x_{i}E_{i}$,
define the projection $\{X>0\}$ by
$\{X>0\}=\sum_{i:x_{i}>0}E_{i}$.
%\begin{equation}
%\{X>0\}=\sum_{i:x_{i}>0}E_{i}.
%\end{equation}
%Similarly,
%we define $\{X<0\}$, $\{X=0\}$ etc.
%to denote the corresponding projections.
Then the trace distance
between quantum states $\rho$ and $\sigma$,
$\tdis(\rho,\sigma)$,
is given by
$\tdis(\rho,\sigma)=\frac{1}{2}\tr|\dr|
=\frac{1}{2}\tr(\dr\{\dr>0\}-\dr\{-\dr>0\})$
%\begin{align}
%\tdis(\rho,\sigma)
%&=\frac{1}{2}\tr|\dr|
%=\frac{1}{2}\tr(\dr\{\dr>0\}-\dr\{-\dr>0\})
%\end{align}
with $\dr=\rho-\sigma$.
The trace distance can be bounded
by another distance called the fidelity as
$\tdis(\rho,\sigma)\le\sqrt{1-\fid(\rho,\sigma)^{2}}$,
%\begin{align}
%\tdis(\rho,\sigma)\le\sqrt{1-\fid(\rho,\sigma)^{2}},
%\end{align}
where the fidelity $\fid(\rho,\sigma)$
between $\rho$ and $\sigma$ is given by
$\fid(\rho,\sigma)=\tr|\sqrt{\rho}\sqrt{\sigma}|$.
%\begin{align}
%\fid(\rho,\sigma)=\tr|\sqrt{\rho}\sqrt{\sigma}|.
%\end{align}

Let $\dte$ be the output of the measurement by Eve.
Then, without loss of generality,
the probability distribution of the random variables
can be written as
\begin{align}
\nonumber
\pba\upco(\bdta,\bdtb,\dte)
&\dfn\prob(\bdta\upco,\bdtb\upco,\dte
|\bbsa\upca,\bbsb\upca,\bdta\upts,\bdtb\upts,\ds,\ts)\\
\nonumber
&=\pba\upco(\bdta)\tr(\msad\upco\otimes\mse)
\unt(\stad\upco\otimes\ste)\untd.
\end{align}
Here, $\ste$ is the initial state
of an ancilla system $\che$ introduced by Eve,
$\mse$ is the Eve's measurement on the ancilla system,
and $\unt$ is the Eve's unitary operation acting
on the composite system.
(The quantum channel is assumed to be
under Eve's control).
For $\cb\subset\co$ and $\pba$ as above,
let $\pba\upcb$ denote
the marginal distribution
of the random variables
defined on $\cb$.
%\begin{align}
%\pba\upts(\bdta,\bdtb,\dte)
%&=\sum_{\bdta\upky,\bdtb\upky,\dte}
%\pba(\bdta,\bdtb,\dte),\\
%\pba\upky(\bdta,\bdtb,\dte)
%&=\sum_{\bdta\upts,\bdtb\upts,\dte}
%\pba(\bdta,\bdtb,\dte),
%\end{align}
%respectively.

We begin with decomposing $\ostal$
($\bsa,\dta\in\bin$) as
\begin{equation}
\label{decst}
\ostad=\pbadz\ostadz+\pbado\ostado,
\quad\ostadz,\ostado\in\csh,
\end{equation}
where
$\pbadz+\pbado=1$,
$\pbad\pbadz=\pbz$
for a positive constant
$\pbz\le\min\oad\{\pbad\}$,
and $\ostadz$ has a Schatten decomposition
of the form
\begin{equation}
\label{decsch}
\ostadz
=\sum_{\kk\oad}\lambda\oad(\kk\oad)
|\kk\oad\ra\la\kk\oad|.
\end{equation}
%\begin{equation}
%\pbad\pbadz=\pbz,\quad
%\pbadz+\pbado=1
%\end{equation}
%It should be mentioned that
We note that
$\ostad$ always has
a decomposition of the above form %
%%%%%
(where we allow $\ostadz=\ostado$)%
.
%%%%%
Let $\sbd=\{(0,0), (0,1), (1,0), (1,1)\}$.
We now construct a set of pure states,
$\{\hst_{\alpha}\}_{\alpha\in\sbd}$,
such that there exists a physical transformation
from $\{\hst_{\alpha}\}_{\alpha\in\sbd}$
to $\{\st_{\alpha}\brz\}_{\alpha\in\sbd}$.
Let $\mab$ ($\alpha,\beta\in\sbd$) be a mapping
from $\{|\kal\ra\}_{\kal}$ to $\{|\kbe\ra\}_{\kbe}$
with $\maa$ being the identity on $\{|\kal\ra\}_{\kal}$,
%$\mab:\{|\kal\ra\}_{\kal}\rightarrow\{|\kbe\ra\}_{\kbe}$,
%$|\kab\ra=\mab(|\kal\ra)$
and introduce the Gram matrix $\grm$ by writing
\begin{equation}
\nonumber
[\grm]_{\alpha\beta}
=\sum_{\kal}
\sqrt{\lal(\kal)\lbe(\kab)}
\la\kal|\kab\ra
\la\phi_{\kal}|\phi_{\kab}\ra,
\end{equation}
where $|\kab\ra=\mab(|\kal\ra)$
and $|\phi_{\kal}\ra$ is a state
on an ancilla system $\ch_{\phi}$.
Since $\grm\ge0$,
there exists a square matrix $\sqm$ such that
$\grm=\sqm^{\dag}\sqm$.
%\begin{equation}
%\grm=\sqm^{\dag}\sqm
%\end{equation}
Further, since all the diagonal elements of $\grm$
are 1,
we can define a pure state $\hst_{\alpha}$
($\alpha\in\sbd$)
on a 4-dimensional Hilbert space $\chfd$ by
\begin{equation}
\nonumber
\hst_{\alpha}=|\sqm_{\alpha}\ra\la\sqm_{\alpha}|,
\end{equation}
where $\sqm_{\alpha}$ denotes
the $\alpha$-th column of $\sqm$.
It follows from this construction that
there exists a physical transformation
from $\{\hst_{\alpha}\}_{\alpha\in\sbd}$
to $\{\st_{\alpha}\brz\}_{\alpha\in\sbd}$
(see \cite{cjw03}).
Now we introduce an approximation of
$\{\hst_{\alpha}\}_{\alpha\in\sbd}$
which is easier to treat in the security proof.
Let $\chtd$ be a 2-dimensional subspace of $\chfd$,
and $\apst\oad$ ($\bsa,\dta\in\bin$)
be states on $\chtd$ such that
\begin{equation}
\nonumber
\apst\ozz+\apst\ozo
=\apst\ooz+\apst\ooo=\id_{\chtd},
\end{equation}
where, for a Hilbert space $\ch$,
$\id_{\ch}$ denotes the identity on $\ch$.
Note that the decompositions
(\ref{decst}) and (\ref{decsch})
and the choises of
$\mab$, $|\phi_{\kal}\ra$ and $\apst\oad$
are not unique;
they should be determined so that
the distance $\tdis(\apst\oad,\hst\oad)$
%$\dr\oad=\tdis(\apst\oad,\hst\oad)$
will be minimized.
%\begin{equation}
%\dr\oad=\tdis(\apst\oad,\hst\oad)
%\end{equation}
In the case of coherent states
with no phase reference,
$\st_{\alpha}=\sum_{\kk\in\N}(\mu^{\kk}/\kk!)e^{-\mu}
|\kk;\alpha\ra\la\kk;\alpha|$,
for instance,
we can take for $\alpha,\beta\in\sbd$ and $\kk\in\N$,
$\ostalz=\hst_{\alpha}
=\apst_{\alpha}=|1;\alpha\ra\la1;\alpha|$,
$\pbalz=\mu e^{-\mu}$,
$\mab(|\kk;\alpha\ra)=|\kk;\beta\ra$ and
$|\phi_{\kk;\alpha}\ra=|\phi\ra$.

The decomposition (\ref{decst}) allows us
to consider that
the Alice's source generates
$\ostadz$ with probability $\pbadz$
and $\ostado$ with probability $\pbado$.
%and if we define $\phadd$ ($\dd\in\bin$) by
%$\phadd=\pbad\pbadd/\sum_{\dtad}{\prob\oadp\prob\oadp\brd}$,
%then $\phadz=\frac{1}{2}$.
Further, we assume that
Eve is informed of partial information about
each state $\stal$ generated by the Alice's source:
(i) $\stal=\ostadz$ or $\stal=\ostado$ and
(ii) $\stal=\ostzdo$ or $\stal=\ostodo$
when $\stal=\ostado$.
This assumption is advantageous to Eve,
and hence does not reduce
the security of the protocol.
Let $\cl\subset\ky$ be the positions where $\ostadz$
is generated, and $\cm=\ky-\cl$.
We now fix $\cl$ and $\cm$,
and consider the best success probability
to estimate $\bdta\upcm$
from $\stad\brom$ and $\bbsa\upcm$.
Here note that
we can estimate each bit $\dta_{\ii}$
of $\bdta\upcm$ separately
because each state $\st\oadi$ is generated
independently of the other bits
$\{\dta_{\ii'}|\ii'\neq\ii,\ii'\in\cm\}$.
For $\bsa\in\bin$,
let $\{\dsaz,\dsao,\dsap\}$
be a POVM on $\ch$
which is used to discriminate
$\st\oaz\bro$ and $\st\oao\bro$,
and let $\pao$ be the conditional probability %that
%$\stal=\st\oaz\bro$ or $\st\oao\bro$;
defined by $\pao=(\pbaz\pbazo+\pbao\pbaoo)/(\pbaz+\pbao)$.
%i.e. $\pao=(\pbaz\pbazo+\pbao\pbaoo)/(\pbaz+\pbao)$.
%\begin{equation}
%\pao=\frac{\pbaz\pbazo+\pbao\pbaoo}{\pbaz+\pbao}
%\end{equation}
Further,
define for a constant $\dma>0$,
%$\pam$ by
%$\pam\pao(\nka/\nda)=(\pma-\dma)$,
%or equivalently
\begin{align}
\nonumber
\pam&=(\pma-\dma)\frac{\nda}{\nka\pao},%\frac{1}{\pao},
\quad\pma=\frac{\nam}{\naa},\\
\nonumber
\epam&=\exp(-\naa\kld(\pdf(\pma)||\pdf(\pma-\dma))),
\end{align}
where
$\nbd=|\{\ii\in\cb|\bbsa_{\ii}=\dd\}|$
for $\cb\subset\ct$,
$\pdf$ denotes the Bernoulli distribution,
and $\kld(\pp||\qq)$ is
the relative entropy of $\pp$ and $\qq$%
%%%%%
\footnote{In the case where Eve is allowed
to collapse Bob, $\nam$ should be replaced by $\nac$.}%
.
%%%%%
Here let us consider the condition $\evt$ given by
\begin{equation}
\nonumber
\evt:\sum_{\dta,\dta'}
\tr\phado\ostado\dsadp\ge\pam,
\end{equation}
where
$\phadd=\pbad\pbadd/(\pbaz\pbazd+\pbao\pbaod)$
for $\dd\in\bin$.
Then it can be verified that
%\begin{gather}
%\pr_{\ct}[\neg\evt]\le
%\epam=\exp(-\na\kld(\pdf(\pma)||\pdf(\pma-\dma)))
%\end{gather}
$\pr_{\ct}[\neg\evt]\le\epam$,
where the probability $\pr_{\ct}$
is taken over the randomness
in choosing $\ds,\ts,\cl\subset\ct$
(see e.g. \cite{ck81}).
Suppose now that
the condition $\evt$ holds.
Then we have
\begin{equation}
\nonumber
\nam\le\nmp\dfn\max_{\cm}\{\nam|\pam\le1\}.
\end{equation}
%\begin{equation}
%0\le\dsad,\quad
%\dsaz+\dsao\le\id_{\ch}
%\end{equation}
Also,
we can write the best success probability
of the discrimination as
\begin{equation}
\nonumber
\suam=\sup_{\dsaz,\dsao:\evt}
\bigg\{\frac{\sum_{\dta}\tr\phado\ostado\dsad}%
{\sum_{\dta,\dta'}\tr\phado\ostado\dsadp}\bigg\}.
\end{equation}
%\begin{equation}
%\phadd=\frac{\pbad\pbadd}{\pbaz\pbazd+\pbao\pbaod},
%\end{equation}
Let $\dtes$ be a random variable
induced by a measurement on $\stad\upky$.
Then, by definition of $\suam$,
it follows that
\begin{equation}
\label{cpres}
\pba\upky(\bdta|\dtes)
\le\pba\upcl(\bdta|\dtes)(\suzm)^{\nzm}(\suom)^{\nom}.
\end{equation}
%\begin{equation}
%\rna\upcm(\bdta|\dte)
%=\ren\upcm(\bdta|\bbsa,\dte)
%=-\nzm\log\suzm-\nom\log\suom
%\end{equation}

Having considered the $\cm$ part,
we next consider the $\cl$ part.
Let us first estimate
the error rate $\ple$ at $\cl$
from $\po={\ner}/{\nt}$,
the error rate at $\ts$.
On remembering that
the error probability of the discrimination at $\cm$
is at least $1-\suam$
for a basis $\bsa$,
define
for a constant $\dps>0$,
\begin{align}
%\label{iepk}
\nonumber
\plmx&=\frac{\nk\po+\nc\dps-\nzm(1-\suzm)-\nom(1-\suom)}{\nl},\\
\nonumber
\epet&=\exp\big(-\nt\kld(\pdf(\po)||\pdf(\po+\dps)\big).
\end{align}
Then we have
$\pr_{\ct}[\ple>\plmx]\le
\mul\dfn\epzm+\epom+\epet$,
%\begin{align}
%%\label{sucpr}
%\pr_{\ct}[\ple>\plmx]\le
%\mul\dfn\epzm+\epom+\epet,
%\end{align}
%i.e.
%\begin{align}
%\kld(p||q)=\sum_{x}p(x)\log\frac{p(x)}{q(x)}
%\end{align}
from which,
it follows that
\begin{equation}
\label{sucpr}
%\nonumber
%\label{condorg}
%\prob\upky(\dtb,\dte|\bbsat)
%=\sum_{\dta\upky:|\dta\oplus\dtb|\le\lceil\nk\pe\rceil}
%\sum_{(\dta,\dtb,\dte)\upky:|\dta\oplus\dtb|>\nk\pkmx}
\sum_{\dta,\dtb,\dte:|\dta\oplus\dtb|>\nl\plmx}
\pba\upcl(\dta,\dtb,\dte)
\le\mul.
\end{equation}
Now, let us consider a modified protocol
in which Alice sends $\hstabd\upcl$
(instead of $\hstad\upcl$),
where $\bbsab$ denotes the bit-wise inversion
of binary string $\bbsa$.
Let $\pbat$ be the corresponding conditional probability
in the modified protocol.
It then follows from the monotonicity of the trace distance
that
\begin{align}
\label{modprt}
\vdis(\pba\upts(\bdta,\bdtbt,\dte),
\pbat\upts(\bdta,\bdtbt,\dte))
\le
\tdis(\hstam\upcl,\hstabm\upcl),
%\dfn\eper
\end{align}
where
$\bst_{\bsa}=\frac{1}{2}\sum_{\dta}\hst\oad$
for $\bsa\in\bin$.
%($\bsa\in\bin$).
%%%%%
We note that $\tdis(\hstam\upcl,\hstabm\upcl)$
can be bounded as
$\tdis(\hstam\upcl,\hstabm\upcl)
\le\sqrt{1-\fid(\hstzm,\hstom)^{2\nl}}$.
%%%%%
%\le(1-\fid(\hstzm,\hstom)^{2\nl})^{1/2}$.
%\begin{equation}
%\tdis(\hstam\upcl,\hstabm\upcl)
%\le\sqrt{1-\fid(\hstzm,\hstom)^{2\nl}}.
%\end{equation}
From inequalities (\ref{sucpr}) and (\ref{modprt}),
it follows that
\begin{equation}
\label{cond}
%\prob\upky(\dtb,\dte|\bbsat)
%=\sum_{\dta\upky:|\dta\oplus\dtb|\le\lceil\nk\pe\rceil}
%\sum_{(\dta,\dtb,\dte)\upky:|\dta\oplus\dtb|>\nk\pkmx}
\sum_{\dta,\dtb,\dte:|\dta\oplus\dtb|>\nl\plmx}
\pbat\upcl(\dta,\dtb,\dte)
\le\mul+\tdis(\hstam\upcl,\hstabm\upcl).
\end{equation}
Let us now introduce the POVM $\{\povmae\}_{\dtb,\dte}$
by writing
\begin{align}
\nonumber
\pba\upcl(\bdta,\bdtb,\dte)
=\tr\hpa\upcl(\bdta)\hstad\upcl\povmae
\end{align}
with
$\hpa\upcl(\bdta)=\prod_{\ii\in\cl}\prh\oadi\brz=2^{-\nl}$,
where, for simplicity,
we have omitted deviding the right-hand side by
$\sum_{\bdtb,\dte}\tr\hstam\upcl\povmae$
because it will be canceled
when we will consider the conditional probability
$\hpa\upcl(\bdta|\bdtbt,\dte)$.
Now,
let us consider the case where
Bob uses the opposite basis $\bbsab$ at $\cl$
and introduce the notation $\bdtbt$ by writing
\begin{align}
\nonumber
\pba\upcl(\bdta,\bdtbt,\dte)
&=\tr\hpa\upcl(\bdta)\hstad\upcl\povmbae. %,\\
%\pba\upcl(\bdtbt,\dte|\bdta)
%&=\tr\stad\upcl\povmbae.
\end{align}
Note that $\mszz+\mszo=\msoz+\msoo$,
%\begin{equation}
%\sum_{\bdtb}\msad\upcl=
%\sum_{\bdtb}\msabd\upcl
%=\id_{\ch}^{\otimes\nk},
%\end{equation}
and so
%\begin{equation}
$\pba\upcl(\bdta,\dte)
=\sum_{\bdtb}\pba\upcl(\bdta,\bdtb,\dte)
=\sum_{\bdtb}\pba\upcl(\bdta,\bdtbt,\dte)$.
%\end{equation}
That is,
the probability distribution $\pba\upcl(\bdta,\dte)$
is independent of the basis
used for the Bob's measurement.
Thus, in the sequel,
we will consider $\pba(\bdta,\bdtbt,\dte)$
rather than $\pba(\bdta,\bdtb,\dte)$.

To examine the security of the protocol,
it is more convenient to treat $\apst\oad$
than $\hst\oad$.
Thus, define
\begin{align}
\nonumber
\hpa\upcl(\bdta,\bdtbt,\dte)
=\tr\hpa\upcl(\bdta)\apad\upcl\povmbae.
\end{align}
The monotonicity
of the trace distance gives
%\begin{gather}
\begin{equation}
\label{tdisdiff}
\begin{split}
&\vdis(\pba\upcl(\bdta,\bdtbt,\dte),
\hpa\upcl(\bdta,\bdtbt,\dte))
\le\nul,\\
%\nonumber
&\nul\dfn\sum_{\bdta}\hpa\upcl(\bdta)
\tdis(\hstad\upcl,\apad\upcl).
\end{split}
\end{equation}
%\end{gather}
This,
together with (\ref{cond}),
yields
\begin{equation}
\label{restmes}
\sum_{\dtb,\dte}\tr(\apabm\upcl-\apaty)\povmbae
\le\mul+\nul+\tdis(\hstam\upcl,\hstabm\upcl),
\end{equation}
where we have defined
\begin{equation}
\nonumber
\apaty=\sum_{\bdta\upcl:|\dta\oplus\dtb|\le\nl\plmx}
\hpa\upcl(\bdta)\apabd\upcl.
\end{equation}

Inequality (\ref{restmes})
can be seen as a restriction
on Eve's measurement.
To take advantage of this restriction,
we now construct a projection on $\ch^{\otimes\nl}$,
$\praty$, which sufficiently preserves $\apaty$.
For this purpose, let us first consider
the problem of quantum hypothesis testing,
where two hypotheses are,
for fixed base $\bsa\in\bin$, 
$H_{\dtz}:\st=\apst\oaz\in\chtd$
and $H_{\dto}:\st=\apst\oao\in\chtd$.
If 
$\{\prad\}_{\dta\in\bin}$, defined by
\begin{equation}
\nonumber
\prad=\{\oapad-\oapadb>0\},
\end{equation}
is used
as a test
for the hypothesis testing,
then the success probability $\sual$
is given by
\begin{align}
\nonumber
\sual=
\frac{1}{2}(1+\tdis(\oapaz,\oapao)).
\end{align}
Suppose now that we receive a product state $\apad\upcl$
from the Alice's source,
and estimate $\bdta\upcl$
by applying
the above hypothesis testing
to each individual state.
Let $\kk$ be an integer such that $0\le\kk\le\nl$.
If we allow up to $\kk$ errors in the estimation
of $\nl$-bit string $\bdta\upcl$,
then the error probability $\eppj$
(i.e. the probability that
we make more than $\kk$ errors)
can be bounded as
\begin{equation}
\nonumber
\eppj\le
%\Big(2^{\nk}-\frac{1}{\nk}2^{\nk\seb(\drp)}\Big)
\bigg(2^{\nl}
-\frac{2^{\nl\seb(\frac{\kk}{\nl})}}{2\sqrt{\nl}}\bigg)
(\suzl)^{\nbzl}(\suol)^{\nbol}
\bigg(\frac{1-\suml}{\suml}\bigg)^{\kk},
%\big((1-\epstm)^{-1}\epstm\big)^{\nk\drp}
\end{equation}
where $\suml=\min\{\suzl,\suol\}$,
$\nbal=\nl-\nal$,
%$\nn^{\s}_{\ky}=|\{\ii\in\ky|\bsb_{\ii}=\s\}|$
%for $\s\in\{0,1\}$,
and we have used,
for $0\le\kk\le\nn$ and $0\le\qq\le1$,
\begin{equation}
\nonumber
\frac{2^{\nn\seb(\frac{\kk}{\nn})}}{2\sqrt{\nn}}
\le\sum_{\ii=0}^{\kk}
\left(\begin{array}{c}
n\\i
\end{array}\right)
\qq^{\ii}(1-\qq)^{\nn-\ii}
\le2^{\nn\seb(\frac{\kk}{\nn})},
\end{equation}
with $\seb(\prob)=-\prob\log\prob
-(1-\prob)\log(1-\prob)$ (see e.g. \cite{ck81}).
We are now in position to construct $\praty$.
Let $\drp=\frac{\kk}{\nl}$ and $\pas=\plmx+\drp$.
Define the projection $\praty$ on $\ch\upcl$ by
\begin{equation}
\nonumber
\praty=\sum_{\bdta\upcl:|\dta\oplus\dtb|\le\nl\pas}
%\pratd
\bigotimes_{\ii\in\cl}\prj_{\bsab_{\ii}\cmm\dta_{\ii}}.
\end{equation}
Then it can be verified that
$\tr\apaty(\idl-\praty)\le\eppj\tr\apaty$,
%\begin{equation}
%\nonumber
%\tr\apaty(\idl-\praty)\le\eppj\tr\apaty,
%\end{equation}
which shows that $\praty$ is a required projection
(provided that $1-\suml$ is sufficiently small).

Having constructed the projection $\prad$,
we now bound the conditional probability
$\hpa\upcl(\bdta|\bdtbt,\dte)$.
Since
\begin{equation}
\nonumber
\hpa\upcl(\bdtbt,\dte)
=\tr\apam\upcl\povmbe=\pilm
\dfn2^{-\nl}\tr\povmbe
\end{equation}
with $\povmbe=\povmbae$ for short,
we now bound 
$\hpa\upcl(\bdta,\bdtbt,\dte)$.
It follows,
on using $\tr\praty\le2^{\nl\seb(\pas)}$, that
\begin{align}
\nonumber
&\tr\praty\pba\upcl(\bdta)\stad\upcl\praty\povmbe
%&\le\tr\praty\stad\upcl\praty
%\nonumber\\
\le\pil,\\
\nonumber
&\pil\dfn2^{-\nl+\nl\seb(\pas)+\nbzl\log\qz+\nbol\log\qo}\tr\povmbe,
\end{align}
where, for $\bsa\in\bin$,
$\qa=\max_{\dta,\dtad\in\bin}
\{\tr\oapad\prj_{\bar{\bsa}\cmm\dtad}\}.$
%\begin{equation}
%\nonumber
%%\label{defqa}
%\qa=\max_{\dta,\dtad\in\bin}
%\{\tr\oapad\prj_{\bar{\bsa}\cmm\dtad}\}.
%\end{equation}
Define now
\begin{equation}
\nonumber
\dpr(\bdta,\bdtbt,\dte)
=\tr(\idl-\praty)\hpa\upcl(\bdta)\apad\upcl(\idl-\praty)\povmbe.
\end{equation}
Since
%$\apam\upcl=(\apabm\upcl-\apaty)+\apaty$,
%$\praty\apaty=\apaty\praty$
$\apam\upcl=\apabm\upcl=(\apabm\upcl-\apaty)+\apaty$,
$\praty$ and $\apaty$ commute,
%\begin{equation}
%\apam\upcl
%=\apabm\upcl
%=(\apabm\upcl-\apaty)+\apaty,
%\end{equation}
%together with
and
$\sum_{\bdtb}\tr\apaty\le2^{\nl\seb(\plmx)}$,
we have
%$\sum_{\bdta,\bdtb,\dte}\hpa\upcl(\bdta,\bdtbt,\dte)
%(\dpr(\bdta,\bdtbt,\dte)/\hpa\upcl(\bdta,\bdtbt,\dte))
%\le\oml$,
\begin{align}
\nonumber
&\sum_{\bdta,\bdtb,\dte}\dpr(\bdta,\bdtbt,\dte)
=\sum_{\bdta,\bdtb,\dte}\hpa\upcl(\bdta,\bdtbt,\dte)
\frac{\dpr(\bdta,\bdtbt,\dte)}{\hpa\upcl(\bdta,\bdtbt,\dte)}
\le\oml,\\
\nonumber
&\oml\dfn\mul+\nul+\tdis(\hstam\upcl,\hstabm\upcl)
+2^{\nl\seb(\plmx)}\eppj.
\end{align}
%where
%\begin{align}
%\nonumber
%\oml=\mul+\nul+\tdis(\hstam\upcl,\hstabm\upcl)
%+2^{\nl\seb(\plmx)}\eppj.
%\end{align}
%That is,
%the expectation of
%${\dpr(\bdta,\bdtbt,\dte)}/{\hpa\upcl(\bdta,\bdtbt,\dte)}$
%over the probability distribution $\hpa\upcl(\bdta,\bdtbt,\dte)$
%is bounded by $\oml$.
Hence
Markov's inequality for a constant $\cns>0$
yields
%%%%%
\begin{equation}
\nonumber
%\label{mar}
\pr_{\hpa}[\dpr(\bdta,\bdtbt,\dte)\le\cns
\oml\hpa\upcl(\bdta,\bdtbt,\dte)]
\ge1-\cns^{-1},
\end{equation}
%%%%%
%$\pr_{\hpa}[\dpr(\bdta,\bdtbt,\dte)>\cns
%\oml\hpa\upcl(\bdta,\bdtbt,\dte)]
%\le\cns^{-1}$, %
%%%%%
where $\cns$ should be determined
so that Eve's mutual information
about the final key
will be minimized%
.
%%%%%
Further, Schwarz's inequality gives
%%%%%
\begin{align}
\nonumber
%\label{sch}
\tr\praty\hpa\upcl(\bdta)\apad\upcl(\idl-\praty)\povmae
&\le(\pil\dpr(\bdta,\bdtbt,\dte))^{\frac{1}{2}}.
%\tr(\idk-\praty)\stad\upcl\praty\povmae
%&\le(\pil\prhd(\bdtbt,\dte|\bbsa,\bdta))^{\frac{1}{2}},
\end{align}
%%%%%
%$\tr\praty\hpa\upcl(\bdta)\apad\upcl(\idl-\praty)\povmae
%\le(\pil\dpr(\bdta,\bdtbt,\dte))^{\frac{1}{2}}$.
%%%%%
Therefore it follows that
%%%%%
\begin{align}
\nonumber
%\label{upbpr}
\hpa\upcl(\bdta,\bdtbt,\dte)
%&=\tr\pba\upcl(\bdta)\stad\upcl\povmae
%\nonumber\\
&\le\big((\pil)^{\frac{1}{2}}
+(\dpr(\bdta,\bdtbt,\dte))^{\frac{1}{2}}\big)^{2},
\end{align}
%%%%%
%$\hpa\upcl(\bdta,\bdtbt,\dte)
%\le\big((\pil)^{\frac{1}{2}}
%+(\dpr(\bdta,\bdtbt,\dte))^{\frac{1}{2}}\big)^{2}$,
%%%%%
and so
\begin{equation}
%\nonumber
\label{defpil}
\pr_{\hpa}[\hpa\upcl(\bdta|\bdtbt,\dte)
>\cpl]\le\frac{1}{\cns},
\quad
\cpl\dfn\frac{\pil}%
{\pilm\big(1-(\cns\oml)^{\frac{1}{2}}\big)^{2}}.
\end{equation}
Now,
it follows from inequality
(\ref{cpres})
that
the conditional {\reny} entropy
$\rna\upky(\rbdta|\bdtbt\upcl,\dte)$
can be bounded as
\begin{align}
\nonumber
\rna\upky(\rbdta|\bdtbt\upcl,\dte)
&\dfn
-\log\sum_{\bdta\upky}
\big(
\pba\upky(\rbdta=\bdta|\rbdtbt=\bdtbt\upcl,\rdte=\dte)
\big)^{2}\\
\nonumber
&\ge\rna\upcl(\rbdta|\bdtbt,\dte)
+\rnamp,
%\nonumber
%\rnaub\upcl(\bdta|\bdtbt,\dte)&=-\log\cpl,
%\rnaub\upcm(\bdta|\dtes)=-\nzm\log\suzm-\nom\log\suom.
\end{align}
where $\rnamp=-\sum_{\bsa}\nam\log\suam$,
and a capital letter (say $\rbdta$)
denotes the random variable
which samples the corresponding small letter (say $\bdta$).
Now,
using constraints (\ref{tdisdiff}) and (\ref{defpil}),
let us derive another constraint of the form
\begin{equation}
\nonumber
\pr_{\pba}[\rna\upcl(\rbdta|\bdtbt,\dte)>\rnalp]
\le\epl.
\end{equation}
If $\nul=0$, for example,
we can take $\rnalp=-\log\cpl$ and $\epl=\cns^{-1}$.
%On remembering inequalities
%(\ref{tdisdiff}) and (\ref{defpil})
%with $\sum_{\dta}\prob(\dta)^{2}\le\max_{\dta}\prob(\dta)$,
%let us define
%\begin{align}
%\nonumber
%\rne&=\min_{\cm:\nam\le\nmp}
%\{-\log\cpl+\rnamp\},\\
%\nonumber
%\epl&=\frac{\nul}{1-\cpl}+\frac{1}{\cns}.
%%\nonumber
%%\rnaub\upcl(\bdta|\bdtbt,\dte)&=-\log\cpl,
%%\rnaub\upcm(\bdta|\dtes)=-\nzm\log\suzm-\nom\log\suom.
%\end{align}
%We can now proceed to bound the {\reny} entropy
%$\ren(\bdtbt\upcl,\dte|\bdta\upcl)$.
Define
%$\rne=\min_{\cm:\nam\le\nmp}
%\{\rnalp+\rnamp\}$,
\begin{align}
\nonumber
\rne=\min_{\cm:\nam\le\nmp}
\{\rnalp+\rnamp\},
\end{align}
and let $\mm$ be an integer
such that $\el\dfn\rne-\mm>0$.
Choose a function $\g$ at random
from a universal family of hash functions
from $\bin^{\nn}$ to $\bin^{\mm}$.
If Alice and Bob choose $\sk=\g(\bdta\upky)$ as their
secret key,
then the Eve's expected information about $\rsk$,
given $\rdte$ and $\rg$, satisfies
$\minf(\rsk:\rdte,\rg)\le\nl\epl+2^{-\el}/\ln2$,
where
we consider $\rbdtbt$ as
an auxiliary random variable
(see \cite{bbcm95} for details).
Here we note that $\rne$ is not explicitly dependent
on the characteristics of the detector,
and hence the detector can be uncharacterized.
Further,
as $\nl\rightarrow\infty$,
the terms $\nul$ and $\tdis(\hstam\upcl,\hstabm\upcl)$
approach to 1
unless $\hst\oad=\apst\oad$ and $\hstam\upcl=\hstabm\upcl$;
this shows that the leading factors
reducing the key generation rate
are the asymmetries of the source
represented by these terms.

To see that our result is consistent
with the previous ones,
suppose that
the source and detector are perfect.
In this case,
we can take $\ostadz=\apad=\ostad$, $\cl=\ky$,
$\mul=\epet$, $\nul=0$, $\tdis(\hstam\upcl,\hstabm\upcl)=0$,
$\log\qa=-1$, $\drp=0$, $\eppj=0$.
Since $\oml=\epet\rightarrow0$
as $\nk\rightarrow\infty$
for fixed $\dps$,
$\rne/\nk$ approaches to $\seb(\po)$
for sufficiently small $\cns^{-1}$
and $\dps$.
This is consistent with
the results in the previous works\cite{gllp04,kp03,ma01,sp00}%
%%%%%
\footnote{%
The original bound given by Mayers\cite{ma01}
is slightly weaker (but this can be improved
by a minor modification).}%.
%%%%%
.

We close this paper with mentioning
some extensions of this work.
(i) In the same way as Koashi-Preskill\cite{kp03},
we can provide a security proof of the BB84 protocol
where the only assumption is that
the detector and basis dependence
of the averaged states %(in terms of the trace distance)
are characterized.
(ii) It is also of importance
to give a security proof of
the B92 protocol\cite{be92}.
Suppose that the source generates
$\stz$ with probability $\praz$
and $\sto$ with probability $\prao$.
Then we decompose $\sta$ ($\bsa\in\bin$) as
$\sta=\prabz\stabz+\prabo\stabo$
so that $\praz\prabz=\prao\prabo$. % and
%$\tdis(\przbo\stzbo,\probo\stobo)$
%will be sufficiently small.
Again we define $\hsta$ by introducing
the Gram matrix as above.
Note that $\hsta$ is a pure state
on a 2-dimensional Hilbert space $\chtd$.
Hence,
the terms
$\nul$ and $\tdis(\hstam\upcl,\hstabm\upcl)$
automatically vanish in this case,
which could be considered as an advantage
of the B92 protocol.
More detailed investigation concerning these extensions
will be the subject of future work.

The author is grateful to Dr. Keiji Matsumoto
for comments.
This work was supported in part
by MEXT,
Grant-in-Aid for Encouragement of Young Scientists (B)
No. 15760289.


\begin{thebibliography}{99}
\bibitem{be92}
C.~H. Bennett,
\prl {\bf 68}, 3121 (1992)
\bibitem{bb84}
C.~H. Bennett and G. Brassard,
in {\em Proceedings of IEEE Conference on Computers,
Systems and Signal Processing},
Banglore (India), 175--179 (1984)
\bibitem{bbcm95}
C.~H. Bennett et al.,
%C. H., Brassard, G., Cr$\rm{\acute{e}}$peau, C.
%and Maurer, U. M.
IEEE Trans. Inform. Theory
{\bf 41}, 1915 (1995)
%\bibitem{bbr88}
%Bennett, C. H., Brassard, G. and Robert J.
%1988 Privacy Amplification by Public Discussion,
%{\em SIAM Journal on Computing}
%{\bf 17}, 210--229
%\bibitem{bgt93}
%Berrou, G., Glavieux, A. and Thitimajshima, P.
%1993 Near Shannon limit error-correcting coding: Turbo codes,
%In {\it Proceedings of IEEE International Conference
%on Communications ICC'93}, Geneva (Switzerland), 1064--1070
%\bibitem{bs93}
%Brassard, G. and Salvail, L.
%1993 Secret key reconciliation by public discussion,
%In {\em Advances in Cryptology - EUROCRYPT'93},
%Lecture Notes in Computer Science {\bf 765}, 410--423
%\bibitem{cm97}
%Cachin, C. and Maurer, U. M.
%1997 Linking Information Reconciliation and
%Privacy Amplification,
%{\em Journal of Cryptology} {\bf 10}, 97--110
%\bibitem{cru01}
%Chung, S.-Y., Richardson, T. J. and Urbanke, R.
%2001 Analysis of Sum-Product Decoding of Low-Density Parity-Check Codes
%Using a Gaussian Approximation,
%{\em IEEE Transactions on Information Theory}
%{\bf 47}, 657--670
\bibitem{cjw03}
A. Chefles, R. Jozsa and A. Winter,
quant-ph/0307227.
\bibitem{ck81}
I. Csiz\'{a}r and J. K\"{o}rner,
Information theory, coding theorems
for discrete memoryless systems,
%Academic Press (1981)
Academic (1981)
\bibitem{gllp04}
D. Gottesman et al.,
%D., Lo, H.-K., Lutkenhaus, N. and Preskill, J.
{\em Quantum Information and Computation} {\bf 4}, 325--360 (2004)
\bibitem{hm03}
M. Hamada,
%{\it Journal of Physics A: Mathematical and General}
J. Phys. A: Math. and Gen.
{\bf 37}, 8303 (2003) 
\bibitem{ha05}
M. Hayashi,
An Introduction to Quantum Information,
Springer, to be published
%\bibitem{hkn00}
%Hirano, H., Konishi, T. and Namiki, R.
%2000 Quantum cryptography using balanced homodyne detection,
%quant-ph/0008037
%\bibitem{klp68}
%Kasami, T, Lin, S. and Peterson, W. W.
%1969 Polynomial Codes,
%{\em IEEE Transactions on Information Theory}
%{\bf 14}, 807--814
%\bibitem{klf01}
%Kou, Y., Lin, S. and Fossorier, M. P. C.
%2001 Low Density Parity Check Codes Based on Finite Geometries:
%A Rediscovery and New Results,
%{\em IEEE Transactions on Information Theory}
%{\bf 47}, 2711--2736
%\bibitem{kf98}
%Kschischang, F. R. and Fray, B. J.
%1998 Iterative Decoding of Compound Codes by
%Probability Propagation in Graphical Models,
%{\it IEEE Journal on Selected Areas in Communications},
%{\bf 16}, 219--230
%\bibitem{lc83}
%Lin, S. and Costello, D. J.
%1983 Error Control Coding: Fundamentals and Applications,
%(Prentice Hall, Englewood Cliffs, New Jersey)
%\bibitem{ma99}
%MacKay, D. J. C.
%1999 Good Error-Correcting Codes Based on Very Sparse Matrices,
%{\em IEEE Transactions on Information Theory}
%{\bf 45}, 399--431
\bibitem{kp03}
M. Koashi and J. Preskill,
\prl {\bf 90}, 057902 (2003)
%\bibitem{lo03}
%Lo, H.-K. 
%2003 Method for decoupling error correction
%from privacy amplification,
%{\em New Journal of Physics} {\bf 5} 36.1--36.24
%\bibitem{mn96}
%MacKay, D. J. C. and Neal, R. M.
%1996 Near Shannon Limit Performance
%of Low Density Parity Check Codes,
%{\it Electronics Letters} {\bf 32}, 1645--1646
%\bibitem{mi02}
%Matsumoto, W. and Imai, H.
%2002 Irregular Extended Euclidean Geometry
%Low-Density Parity-Check codes,
%In {\em Proceedings of Communication Systems, Networks
%and Digital Signal Processing CSNDSP 2002},
%Staffordshire University (UK), 148--151
%\bibitem{mxi02}
%Matsumoto, W., Xu W. and Imai, H.
%2002 Irregular Low-Density Parity-Check Code Design
%based on Euclidean Geometries,
%{\em to appear in IEICE Trans. Fundamentals}
%\bibitem{mmc98}
%McEliece, R. J., MacKay, D. J. C. and Cheng, J.-F.
%1998 Turbo Decoding as an Instance of
%Pearl's ``Belief Propagation'' Algorithm,
%{\it IEEE Journal on Selected Areas in Communications}
%{\bf 16}, 140--152
\bibitem{ma01}
D. Mayers,
J. Assoc. Comput. Mach.
{\bf 48}, 351 (2001)
%\bibitem{nea01}
%Nielsen P. M., Schori, C., Sorensen, J. L.,
%Salvail, L., Damgard, I. and Polzik, E.
%2001 Experimental quantum key distribution
%with proven security against realistic attacks
%{\it Journal of Modern Optics}
%{\bf 48}, 1921--1942
%\bibitem{pe88}
%Pearl, J.
%1988 {\it Probabilistic Reasoning in Intelligent Systems:
%Networks of Plausible Inference}
%(Morgan Kaufmann, San Mateo).
%\bibitem{ru01}
%Richardson, T. J. and Urbanke, R.
%2001 The capacity of low-density parity-check codes
%under message-passing decoding,
%{\em IEEE Transactions on Information Theory}
%{\bf 47}, 599--618
\bibitem{sp00}
P. Shor and J. Preskill,
\prl {\bf 85}, 441 (2000)
\bibitem{wmu04}
%S. Watanabe et al., %, R. Matsumoto and T. Uyematsu,
S. Watanabe, R. Matsumoto and T. Uyematsu,
quant-ph/0412070
\end{thebibliography}
\end{document}